\documentclass[twoside]{article}
\usepackage{Proc_NTSE_14}
\usepackage{sidecap}
\begin{document}
\thispagestyle{plain}
\publref{myfilename}

\begin{center}
{\Large \bf \strut
Computational approaches to many-body
dynamics of unstable nuclear systems
\strut}\\
\vspace{10mm}
{\large \bf 
Alexander Volya
}
\end{center}

\noindent{
\small Department of Physics, Florida State University,
Tallahassee, FL 32306-4350, USA} 

\markboth{A.Volya}
{Computational approaches to many-body
dynamics of unstable nuclear systems} 

\begin{abstract}
The goal of this presentation is to highlight various computational techniques used to study dynamics of quantum many-body systems. 
We examine the projection and variable phase methods being applied to 
multi-channel problems 
of scattering and tunneling; here the virtual, energy-forbidden channels and their treatment are of particular importance. 
The direct time-dependent solutions using Trotter-Suzuki propagator expansion  
provide yet another approach to exploring the complex dynamics of unstable systems. 
While presenting computational tools, we briefly revisit the general theory of the quantum decay of unstable states. 
The list of questions here includes those of the 
internal dynamics in decaying systems, formation and evolution of the radiating state, and 
low-energy background that dominates at remote times. Mathematical formulations and numerical approaches 
to time-dependent problems are discussed using the quasi-stationary methods involving effective 
Non-Hermitian Hamiltonian formulation. 
\\[\baselineskip] 
\end{abstract}

\section{Introduction}
There is no physical system that is truly isolated from the rest of the world, the closed system idealization may be 
convenient but becomes poor or completely invalid for many questions of modern-day science. 
In nuclear physics, as interests shift towards weakly bound, unbound or even dynamically evolving reaction states, 
the theoretical approaches for dealing with unstable dynamics of open quantum systems with multiple degrees of freedom 
must be revisited.  The availability of advanced computational technologies calls forth innovative thinking and new philosophies in 
addressing these types of quantum many-body problems. 
In this presentation, using different models and realistic examples 
from the world of nuclear physics, we discuss computational strategies and techniques for dealing with dynamically 
unstable many-body systems. The {\it Nuclear Theory in the Supercomputing Era} venue is especially timely and 
allows us to put emphasis on some of the techniques, that due to their computational nature, remained behind 
the curtains in a number of recent investigations \cite{Volya:2009,Ahsan:2010,Peshkin:2014}.

\section{Intrinsic degrees of freedom in reactions}
\subsection{Projection method \label{sec:proj}}
\begin{figure}[!ht]
\begin{minipage}[h]{0.5\textwidth}
\includegraphics[width=2.5in]{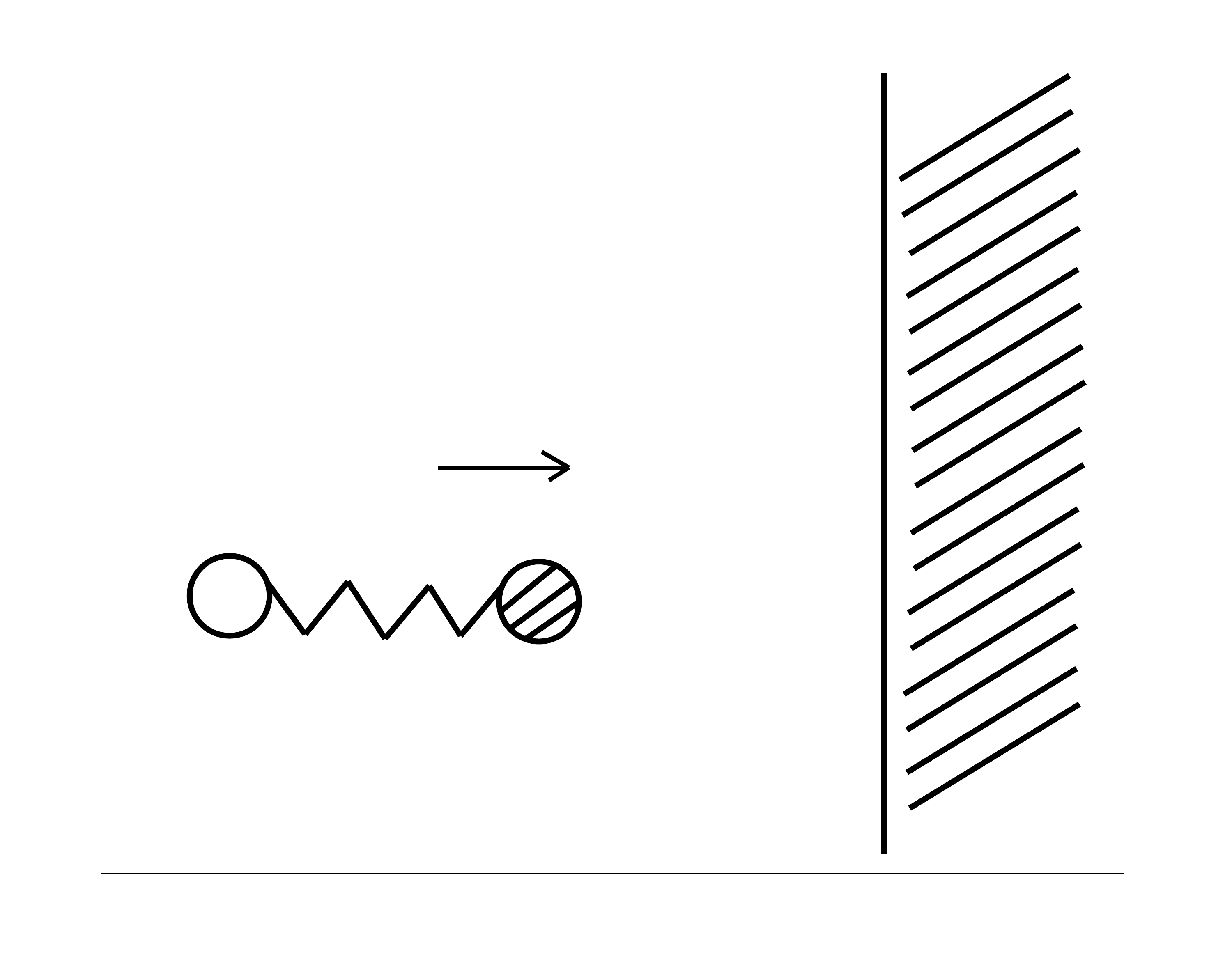} 
\end{minipage}
\begin{minipage}[h]{0.5\textwidth}
\caption{Schematic picture of scattering. A composite system of two particles bound by a harmonic oscillator potential scatters off an
infinite wall. One of the particles does not interact with the wall, at the same time the wall is impenetrable for the second particle. 
\label{fig:Schematic-picture}}
\end{minipage}
\end{figure}

Let us start by illustrating the difficulties that one faces 
while trying to reformulate reaction problems using the basis projection methods typical for structure physics; 
see also Refs.~\cite{Ahsan:2011,Ahsan:2010}. Consider a model of scattering illustrated in Fig.~\ref{fig:Schematic-picture}.
In this one-dimensional problem two particles with masses $\mu_{1}$ and $\mu_{2}$ comprise a composite system of 
unit mass $\mu_{1}+\mu_{2}=1.$ The system can be described with the center-of-mass and relative coordinates, $X=\mu_{1}x_{1}+\mu_{2}x_{2}$
and $x=x_{1}-x_{2}$, respectively. The two particles are confined by a potential $v(x).$ The intrinsic Hamiltonian  
\begin{equation}
h=-\frac{1}{\mu}\frac{\partial^{2}}{\partial x^{2}}+v(x)
\label{eq:Hamiltonianin}
\end{equation}
is assumed to have a complete set of discrete eigenstates $\psi_{n}(x)$ with corresponding intrinsic energies $\epsilon_{n}$:
\[h\psi_{n}(x)=\epsilon_{n}\psi_{n}(x),\qquad n=0,1,2,...\]
Here the reduced mass is $\mu=\mu_{1}\mu_{2}$  and we select our units so that $\hbar^{2}/2=1.$
We assume that this system scatters off an infinite wall and the wall interacts only with the second particle. Therefore the
full Hamiltonian is 
\begin{equation}
H=-\frac{\partial^{2}}{\partial X^{2}}+U(x_{2})+h,\,\,{\rm where}\quad
U(x_{2})=\left \{ 
\begin{array}{cc}
\infty & {\rm if}\quad x_2\ge 0 \cr
0 &{\rm if}\quad x_2< 0
\end{array}
\right . .
\label{eq:Hamiltonian}
\end{equation}
As illustrated in Fig.~\ref{fig:Schematic-picture}, we assume 
that the incident beam is traveling from the left
and contains the projectiles in an intrinsic state (channel) $n.$ 
A complete set of reflected waves is characterized by the amplitudes $R_{nm}$ defined here so that  $|R_{mn}|^{2}$ 
represents the probability for the initial beam  in channel $n$ to reflect in channel
$m;$ $R_{nm}=R_{mn}$ due to time-reversal invariance. 
The
scattering wave function is 
\noindent \begin{equation}
\Phi(X,x)=\frac{e^{iK_{n}X}}{\sqrt{|K_{n}|}}\psi_{n}(x)+\sum_{m=0}^{\infty}\frac{R_{mn}}{\sqrt{|K_{m}|}}e^{-iK_{m}X}\psi_{m}(x),
\label{eq:wallWF-1}
\end{equation}
\begin{equation} {\rm{where}} \qquad K_{n}(E)=\sqrt{(E-\epsilon_{n})}\label{eq:define_K-1} \end{equation}
is the center-of-mass momentum of the
two-particle system while in the $n^{th}$ intrinsic state, and $E$ is the total energy.

A channel $n$ is considered to be open if $E\ge\epsilon_{n}$ and the corresponding momentum $K_{n}$ is real. 
The conservation of particle-number in all the open channels necessitates  $\sum_{m\in{\rm open}}{|R_{mn}|^{2}}=1.$ 
The channel is closed if $E<\epsilon_{n},$ in which case $K_{n}$ is purely imaginary. We stress that the principal value of the 
square root is implied in Eq. (\ref{eq:define_K-1}).

The boundary condition set by an impenetrable wall 
\begin{equation}  \Phi(X,x)=0\,\,\,{\rm at}\,\, x_{2}=0\label{eq:boundary}  \end{equation}
is to be used for determining the set of coefficients $R_{mn}.$ Since at $x_{2}=0$ the center-of-mass coordinate $X=\mu_{1}x,$
the boundary condition can be expressed in the intrinsic coordinate $x$ only $\Phi(\mu_{1}x,\, x)=0.$
Therefore we can project the reaction problem onto a complete set of intrinsic basis states, 
which leads to the following linear equation
\begin{equation}
\sum_{m}\frac{D_{n'm}\left[-i\mu_{1}({K}_{n}+{K}_{m})\right]}{\sqrt{|K_{m}|}}\, R_{mn}=-\frac{\delta_{n'n}}{\sqrt{|K_{n}|}}\,,\label{eq:D-matrix}\end{equation}
where the matrix $D$ is defined as \begin{equation}
D_{mn}(\varkappa)=\langle \psi_{m}|\exp(\varkappa x)|\psi_{n}\rangle\,.\end{equation}
Eq. \eqref{eq:D-matrix} represents a typical mathematical challenge associated with the formulation of reaction problems 
where reaction states are 
projected onto the intrinsic states; see also Sec.~\ref{sec:tdcsm}. 
It is a linear algebra problem where the construction of the scattering matrix 
amounts to matrix inversion in the projected space. 
The scattering energy $E$ is a running parameter here, and studies of scattering at different energies is therefore time consuming. 
And, finally, the underlying matrix is highly singular and there are issues with convergence. 
The latter difficulty is the one that we would like to illustrate using this example. 

\begin{figure}[!ht]
\begin{centering}
\begin{tabular}{cc}
\includegraphics[width=6cm]{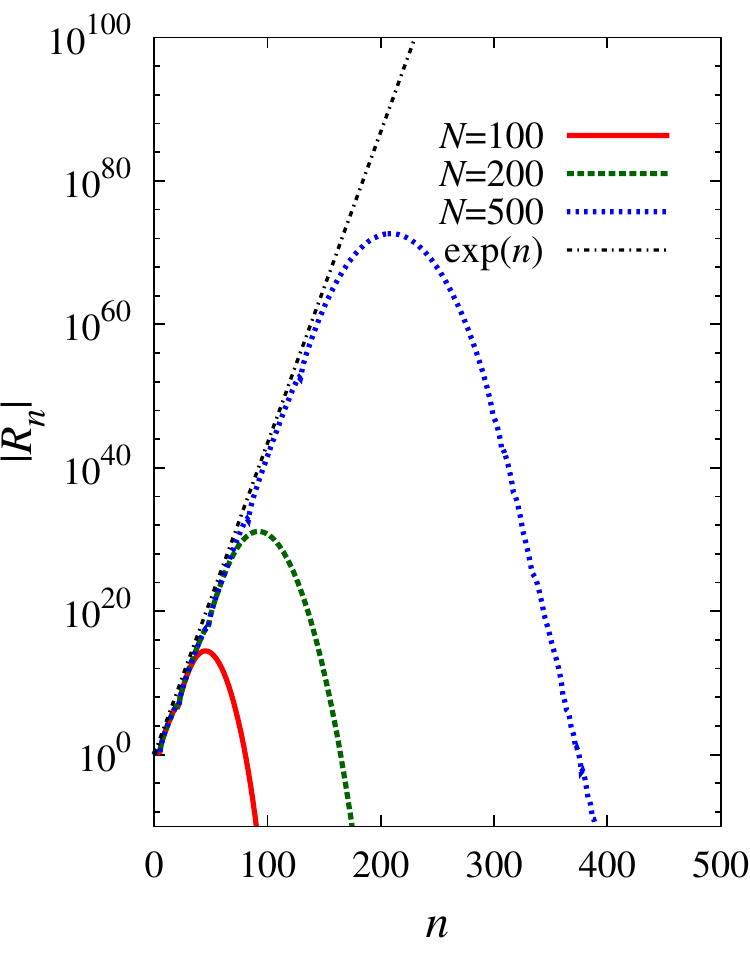} & \includegraphics[width=6.28cm]{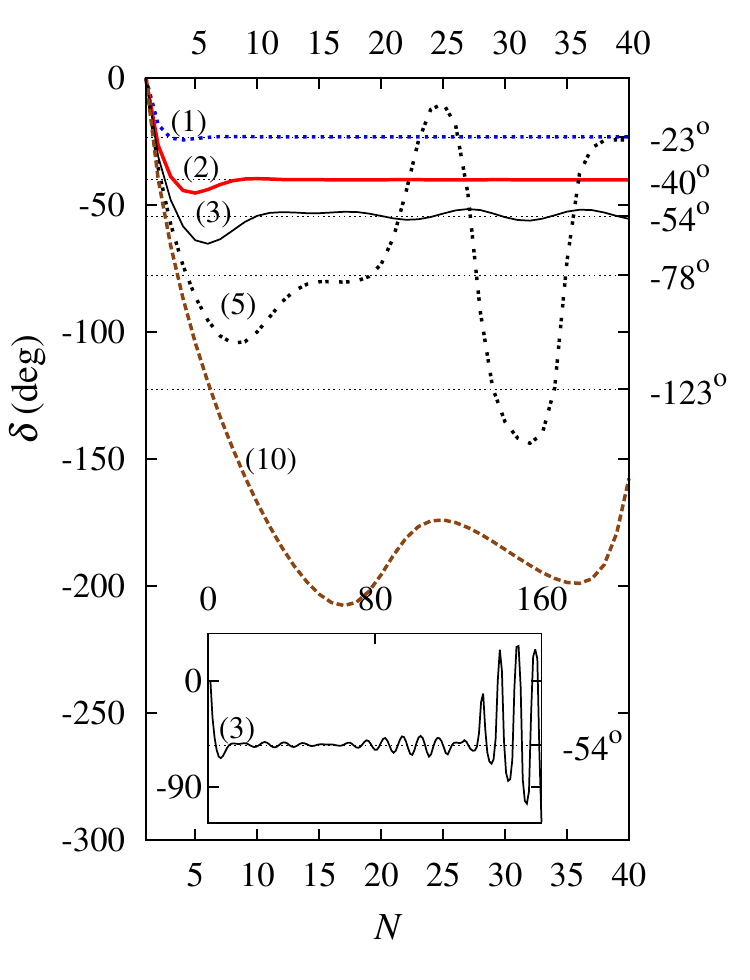} \tabularnewline
\end{tabular}
\par\end{centering}
\caption{This figure refers to a system of two particles, bound by a harmonic oscillator confinement, which collides with an infinite wall. The 
incident kinetic energy is exactly half of the oscillator quantum so that only the ground state channel is open. 
\hspace{\textwidth}
Left panel:  For a system where $\mu_{1}=\mu_{2}$ the absolute values of amplitudes $|R_n|\equiv |R_{0n}|$ in virtual channels 
are shown as a function of $n$ assuming different truncations $N$. The asymptotic dependence is illustrated with the straight line ``$\exp(n)$.''  
\hspace{\textwidth}
Right panel: The phase shift, defined for a single open channel as $e^{2i\delta}=-R_{00},$
is plotted as a function of truncation $N.$ The problems with the approach are highlighted by an unstable and oscillatory behavior of the 
phase shifts. The problem is particularly  severe when the non-interacting particle of mass $\mu_1$ is heavy.
Different curves show phase-shifts for different mass ratios $\mu_1/\mu_2=1, 2, 3, 5, 10$, as labeled; the exact values obtained with 
Variable Phase Method (see Sec.~\ref{sec:VPM})  
are shown by the horizontal grid lines with the tic-marks on the right.
Inset shows the case when $\mu_1/\mu_2=3$ extending the study  to considerably large values of $N$ and emphasizing that for any choice of parameters  the approach fails at some point. 
\label{fig:proj}
}
\end{figure}

If the two particles forming a composite system are bound by a harmonic
oscillator confinement, $v(x)=\mu\omega^2 x^2/2$ in Eq.~(\ref{eq:Hamiltonianin}), the $D$-matrix is then known analytically \cite{Ahsan:2010}. 
Then to solve the problem we truncate the channel space at some large number $N$ of oscillator quanta, and solve  Eq.~\eqref{eq:D-matrix} 
using standard numerical techniques.
This turns out to be a difficult task; 
the matrix element $D_{mn}(\varkappa)$ for virtual channels, where  $\varkappa$ is real, are exponentially large, making the process of matrix inversion
difficult and numerically unstable \cite{Sakharuk:1999,MORO:2000,Ahsan:2010}. 
As shown in  Fig.~\ref{fig:proj}, left panel, the absolute values of the 
reflection amplitudes, $R_n\equiv R_{0n}$ exponentially diverge for increasingly remote virtual channels.

While it is possible to overcome the numerical issues, further
examination shows that the approach has  fundamental flaws. 
In Fig.~\ref{fig:proj}, right panel, the phase shift, defined as $e^{2i\delta}=-R_{00}$, is shown as a function of  $N$. 
While satisfactory and seemingly convergent results can be easily found for the cases where the mass of the non-interacting particle is small, 
in general, as $N$ increases, the results start oscillating; 
situations where the non-interacting particle is heavy and therefore deeply penetrates the wall are particularly difficult to handle. 
It was emphasized in Refs. \cite{Ahsan:2010,Ahsan:2011} that there is no numerical convergence
with increasing $N$. 

\subsection{Variable Phase Method\label{sec:VPM}}
The above example shows that reaction problems call for new techniques. 
One approach, based on the Variable Phase Method (VPM), see Ref. \cite{Babikov:1968}, is proposed in Ref.~\cite{Ahsan:2010}. 
The VPM is an effective technique for solving the coupled-channel problem of the form 
\begin{equation}
\left[\frac{\partial^{2}}{\partial X^{2}}+K_{n}^{2}\right]\Psi_{n}(X)-\sum_{n'}V_{nn'}(X)\Psi_{n'}(X)=0,\label{eq:cc_SE}
\end{equation}
where scattering observables are to be expressed relative to free-space solutions normalized to unit current
\begin{equation}
\Xi_{nn'}^{\pm}(X)=\frac{e^{\pm iK_{n}X}}{\sqrt{-2iK_{n}}}\,\delta_{nn'};\label{eq:VPM-OPM}\end{equation}
the $\pm$ sign corresponds to a wave moving in the right/left
direction. 
In the VPM approach the coupled-channel Sch\"odinger's equation (\ref{eq:cc_SE}) is reformulated as a set of first order differential equations 
for dynamic reflection and transmission amplitude matrices $R_{nn'}(X')$ and $T_{nn'}(X').$ These amplitudes 
correspond to a potential that is cut at $X',$ namely 
to $V_{nn'}(X)\theta(X-X'):$
\begin{equation}
\frac{dR(X)}{dX}=\left[\left(\Xi^{+}+R(X)\,\Xi^{-}\right)\right]V\left[\Xi^{+}+\Xi^{-}\, R(X)\right],
\quad R_{nn'}(\infty)=0,
\label{eq:VPM_R}\end{equation}
\begin{equation}
\frac{dT(X)}{dX}=T(X)\,\Xi^{-}\, V\,\left[\Xi^{+}+\Xi^{-}\, R(X)\right],\quad T_{nn'}(\infty)=\delta_{nn'}.
\label{eq:VPM_T-1}\end{equation}
These equations, being solved from $X=+\infty$ towards $X\rightarrow -\infty,$ recover the reflection and transmission amplitudes 
$R_{nn'}(-\infty)=R_{nn'}$ and $T_{nn'}(-\infty)=T_{nn'}.$

Using factorization of the form \[\Phi(X,x)=\sum_{n}\Psi_{n}(X)\psi_{n}(x)\] the  
Schr\"odinger's equation for the scattering problem described in Fig.~\ref{fig:Schematic-picture}
can be transformed into a coupled-channel equation (\ref{eq:cc_SE})
for the center-of-mass
wave-functions $\Psi_{n}(X)$ 
where the folded potentials are 
\begin{equation}
V_{nn'}(X)=
\int_{-\infty}^{\infty}\psi_{n}^{*}(x)\, U(X,x)\,\psi_{n'}(x)dx\,.\label{eq:FoldedPot}
\end{equation}

\begin{figure}[!ht]
\begin{minipage}[h]{0.6\textwidth}
\includegraphics[width=0.95\linewidth]{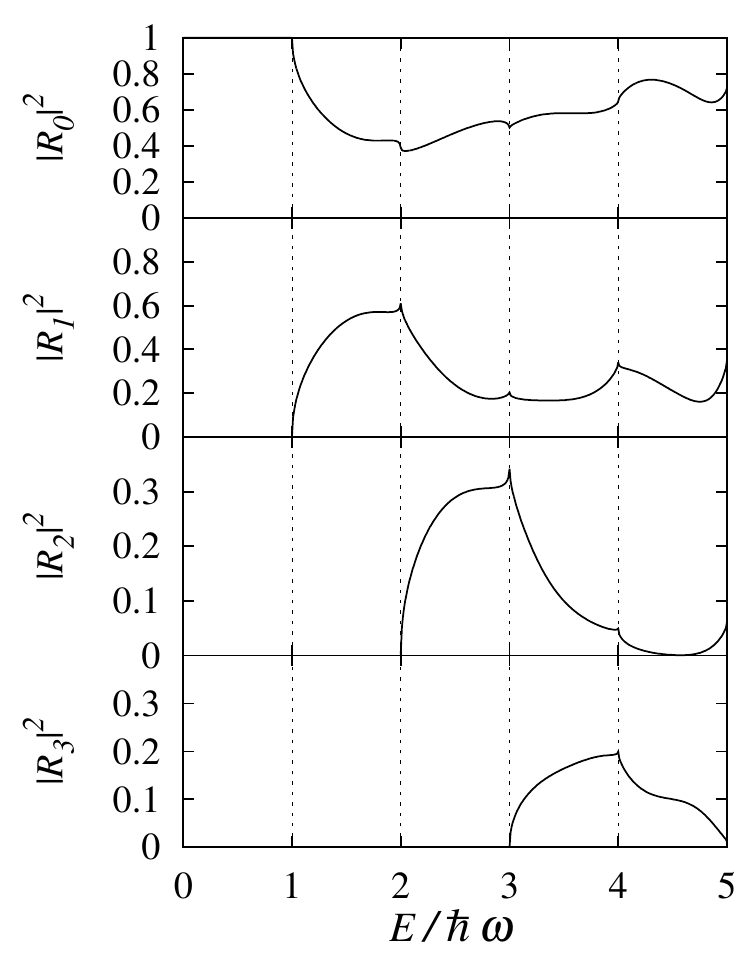}
\end{minipage}
\begin{minipage}[h]{0.55\textwidth}
\caption{
Reflection probabilities in different channels as a function of incident kinetic energy. The incident beam contains a composite projectile 
in the ground state. Equal masses $\mu_1=\mu_2$ are assumed for both interacting and non-interacting particles. 
\label{fig:oscillator_r}
}
\end{minipage}
\end{figure}

\begin{figure}[!ht]
\begin{minipage}[h]{0.6\textwidth}
\includegraphics[width=0.95\linewidth]{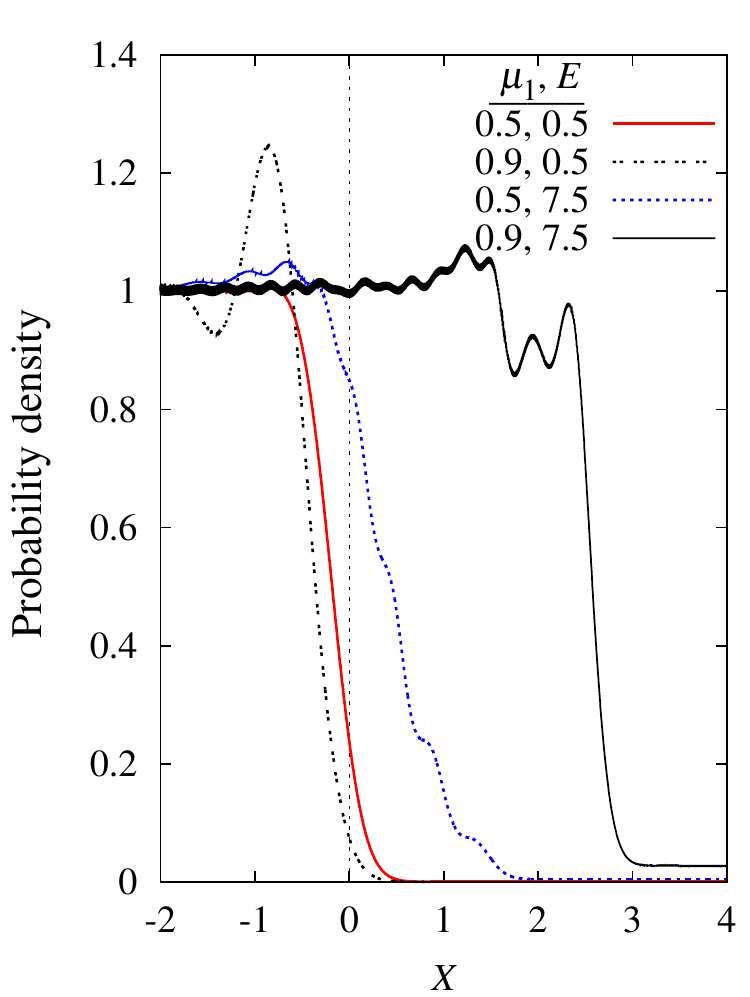}
\end{minipage}
\begin{minipage}[h]{0.55\textwidth}
\caption{
The density of probability for the center of mass of the projectile
to be at a location $X$ when it is reflected from an infinite wall at $X=0.$
\label{fig:osc_wallnorm}
}
\end{minipage}
\end{figure}
Some representative results for the scattering problem where an oscillator-bound system interacts with an infinite wall
are shown in Figs.~\ref{fig:oscillator_r} and \ref{fig:osc_wallnorm}. The reflection probabilities
for different channels
are shown in Fig.~\ref{fig:oscillator_r} as functions
of incident kinetic energy. The kinetic energy is expressed in units of oscillator's $\hbar \omega$ and therefore for each integer
value thereof a new channel opens. One can notice typical cusps at thresholds associated with the loss of flux into newly opened channels. 
In Fig.~\ref{fig:osc_wallnorm} the probability distribution for the center of mass is shown. 
The four curves show four of the most representative situations; low and high incident kinetic 
energies $E=0.5 \hbar\omega$ and $E=7.5 \hbar \omega,$ respectively, and two different mass-ratios $\mu_{1}=0.5$ and 0.9.

\subsection{Time-dependent approach}
Turning to a time-dependent approach is a natural strategy for dealing with non-stationary systems. 
There are various computational techniques; see Ref.~\cite{vanDijk:2011} for some recent tests and comparisons of methods 
being applied to one dimensional Schr\"odinger's equation. 
In time-dependent techniques preservation of unitarity is often at the core of computational difficulties: 
lack of unitarity could lead to exponential
amplification of numerical noise even for single channel, 
while in multi-channel problems discontinuities near thresholds are particularly challenging. 
Here we propose and demonstrate another approach that is computationally efficient, even in multi-variable cases, and
preserves unitarity exactly. 

The time propagation 
\begin{equation}
\Phi({\bf x}, t)=\exp\left(  -\frac{i}{\hbar} H t \right)\Phi({\bf x}, 0)
\label{eq:evolve}
\end{equation}
can be performed by considering, separately, the potential and kinetic parts of the hamiltonian $H=K+V.$ 
In the discretized space of generalized coordinates ${\bf x}=\{x_1, x_2,\dots\}$ the potential $V({\bf x})$ is diagonal, so that the 
exponential operator $\exp(-iV t/\hbar )$ 
can be readily applied. Similarly, in the conjugate momentum space ${\bf p}=\{p_1,p_2\dots\}$ the propagation with kinetic 
energy operator, which is diagonal, is also easy to perform. 
While the operators $K$ and $V$ do not commute, the time evolution (\ref{eq:evolve}) with the
combined Hamiltonian can be done efficiently with the Trotter-Suzuki approach \cite{Trotter:1959t,Suzuki:1990s}. In this approach
the propagation is done in small time steps $\Delta T;$ for each of these steps the evolution operator is approximated as
\begin{equation}
\exp\left(  -\frac{i}{\hbar} H \Delta t \right)=\exp\left(  -\frac{i}{2\hbar} V \Delta t \right) \exp\left(  -\frac{i}{\hbar} K \Delta t \right) \exp\left(  -\frac{i}{2\hbar} V \Delta t \right) + O(\Delta t^3).
\label{eq:ts}
\end{equation}  
The Fast Fourier Transform allows for an efficient transition between coordinate and momentum representations so that exponentials of operators
are always applied in the diagonal form. Even with the finite time steps the unitarity is fully retained; the method is applicable to time-dependent 
Hamiltonians. The computational cost of two back and forth Fourier transforms involved in each step is $N \log(N)$ assuming the coordinate space
is discretized into $N$ points.  While this at first appears to be higher than the typical $O(N)$ scaling of the traditional methods, in practice 
the cost $c N$ of any high quality method involves a constant factor $c$ that often exceeds $\log(N).$   Moreover, modern computer 
hardware often comes with signal processing tools which are optimized at hardware and software level to perform Fast Fourier Transform
with incredible efficiency. 

\begin{figure}[!ht]
\begin{centering}
\begin{tabular}{cc}
\includegraphics[width=0.475\linewidth]{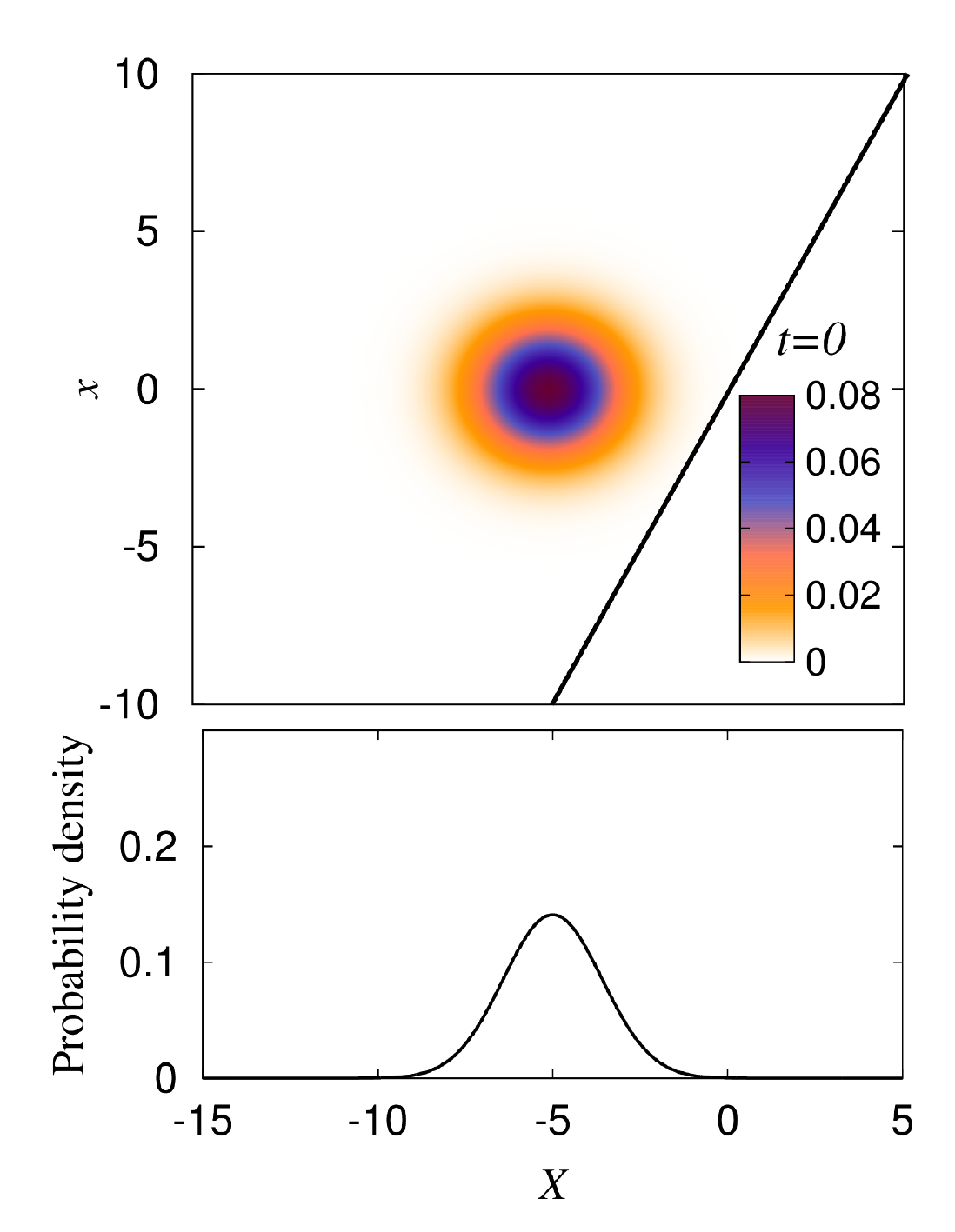} & \includegraphics[width=0.475\linewidth]{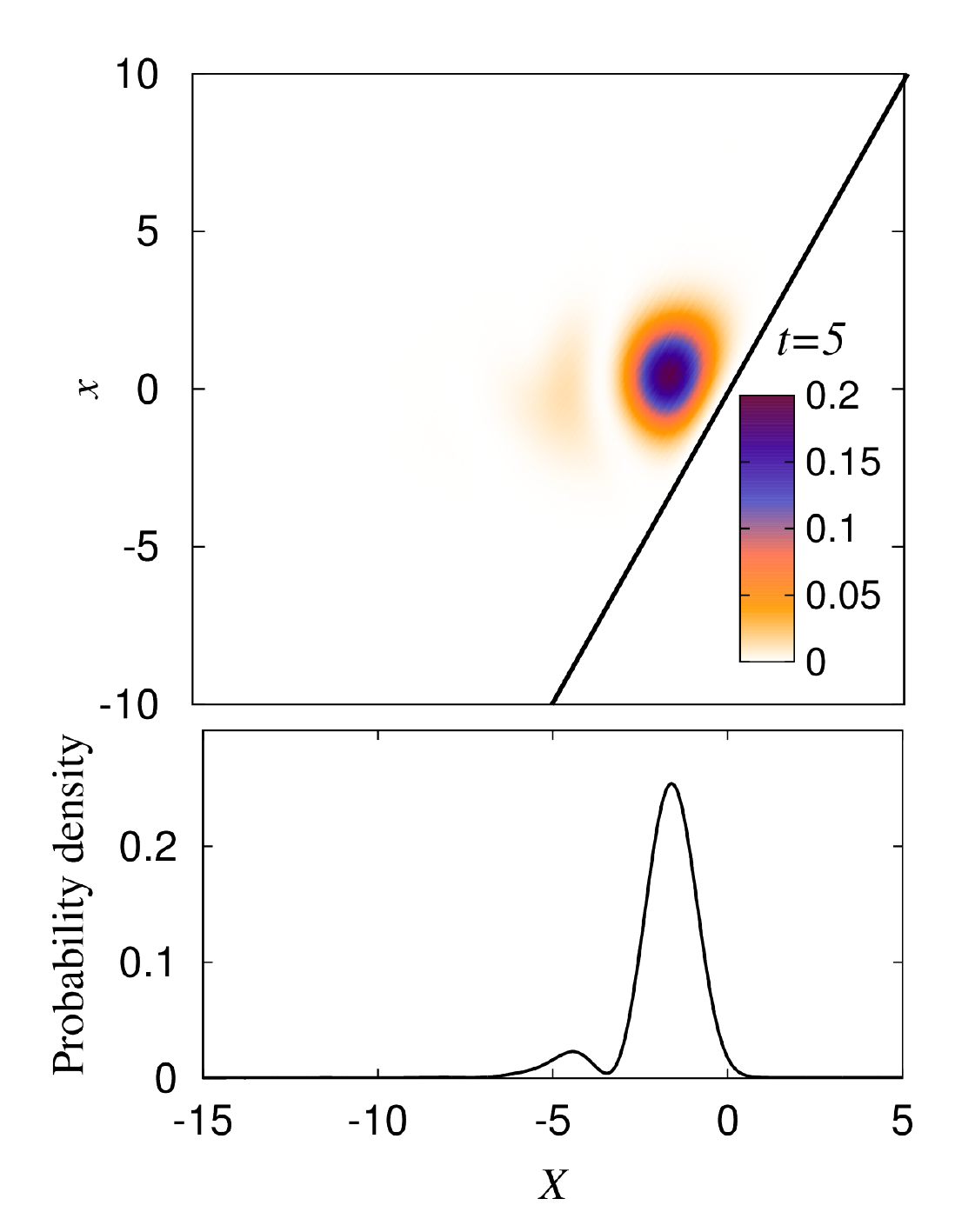} \tabularnewline
\includegraphics[width=0.475\linewidth]{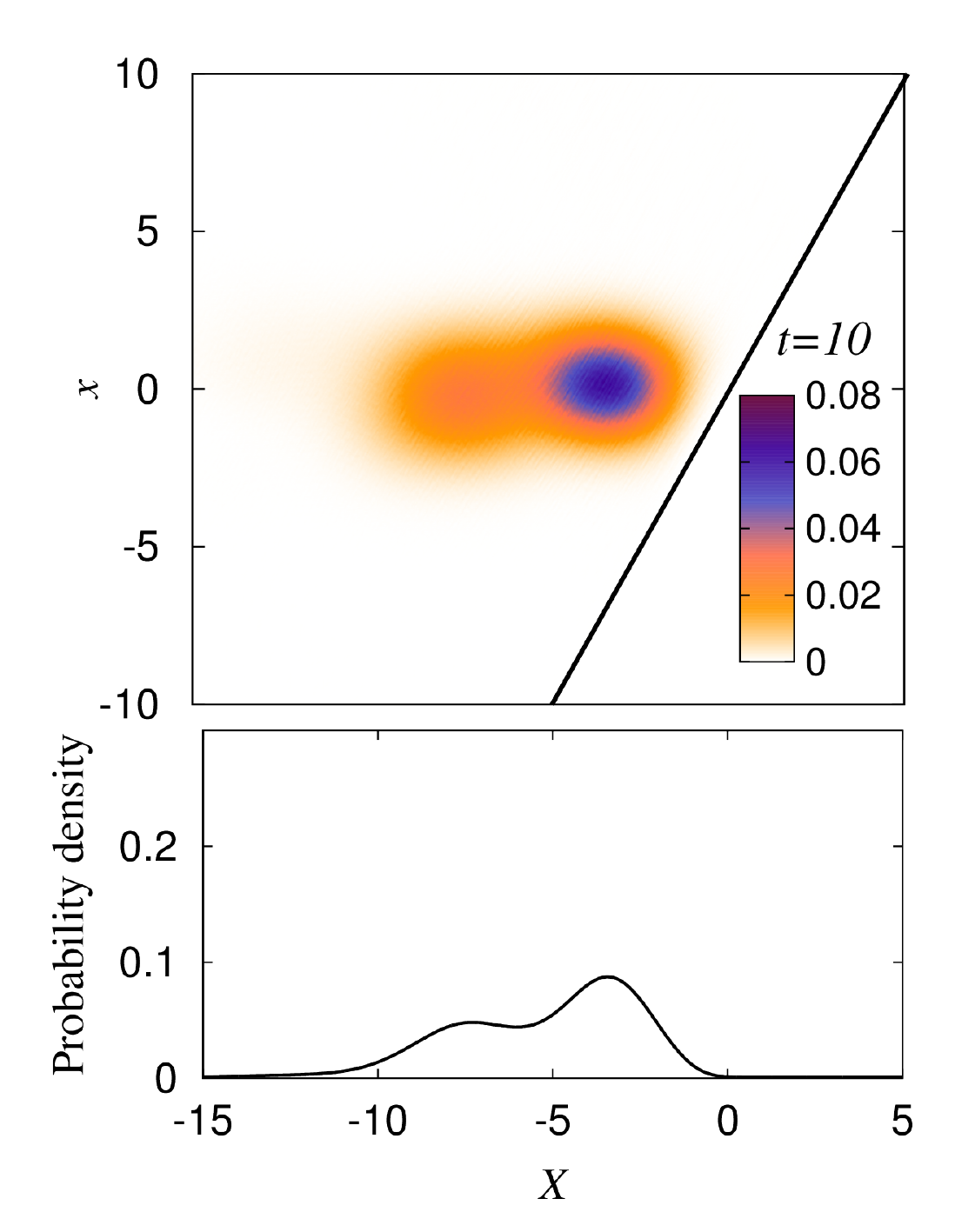} & \includegraphics[width=0.475\linewidth]{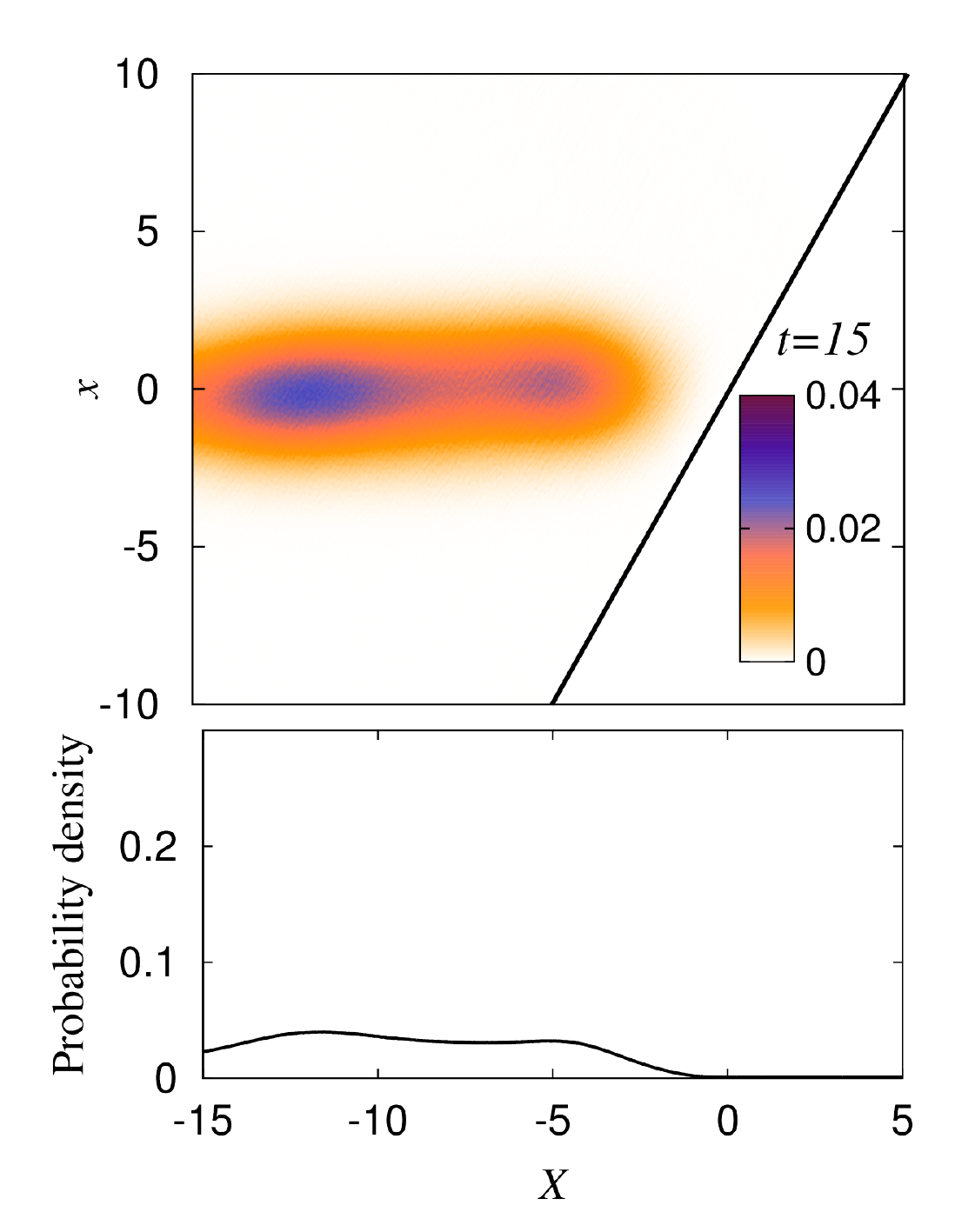} \tabularnewline
\end{tabular}
\end{centering}
\caption{
Four panels show the wave function $|\Phi(X,x)|^2$ as a density plot for different times $t=0, 5,10$ and 15, as labeled. For each of the 
time snapshots the lower plot shows the density distribution over the center of mass coordinate computed as $\int |\Phi(X,x)|^2 dx.$ The initial
wave function at $t=0$ is given the Gaussian wave packet, Eq. \ref{eq:istate}. For this system $\mu_1=\mu_2$, the border of inaccessible area $x_2>0$ is shown with a solid line. 
\label{fig:1dtime}}
\end{figure}

\begin{figure}[!ht]
\begin{tabular}{cc}
\includegraphics[width=0.475\linewidth]{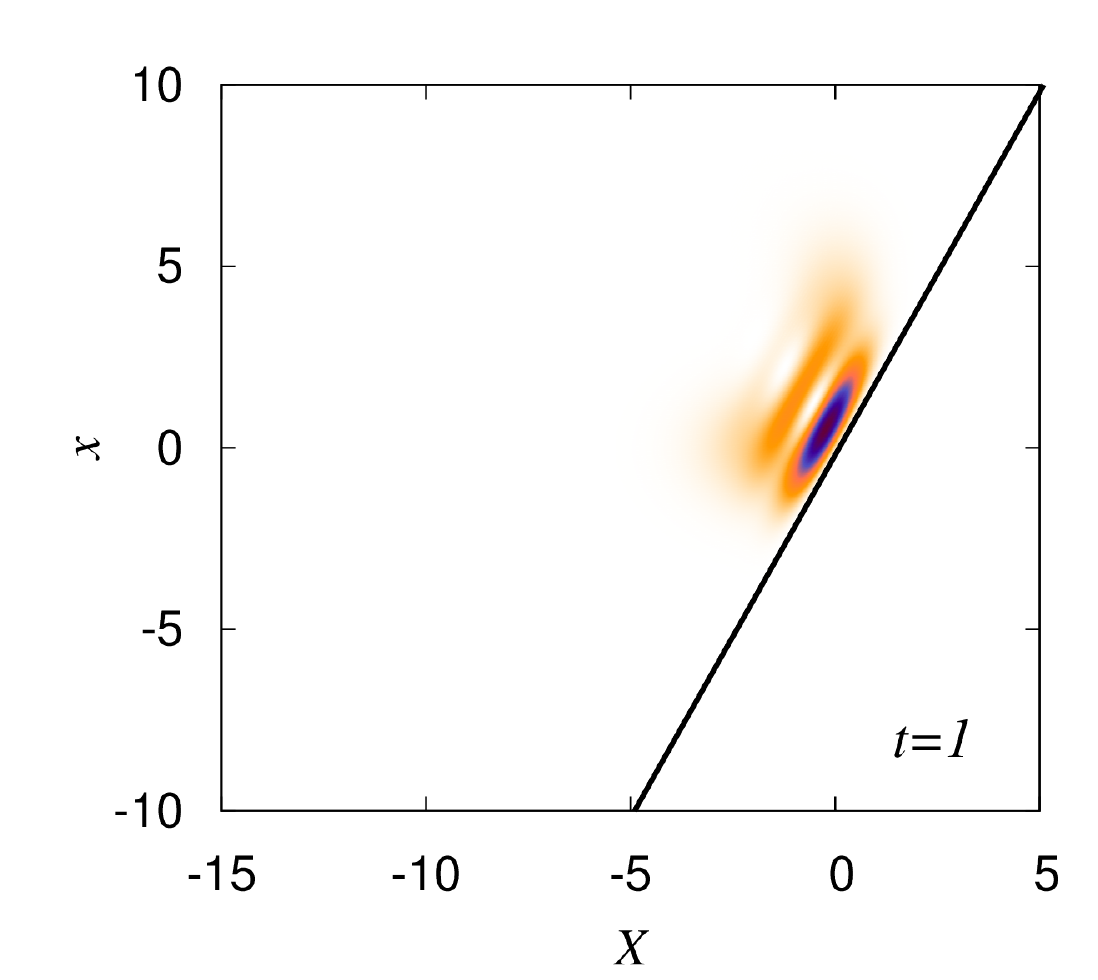} & \includegraphics[width=0.475\linewidth]{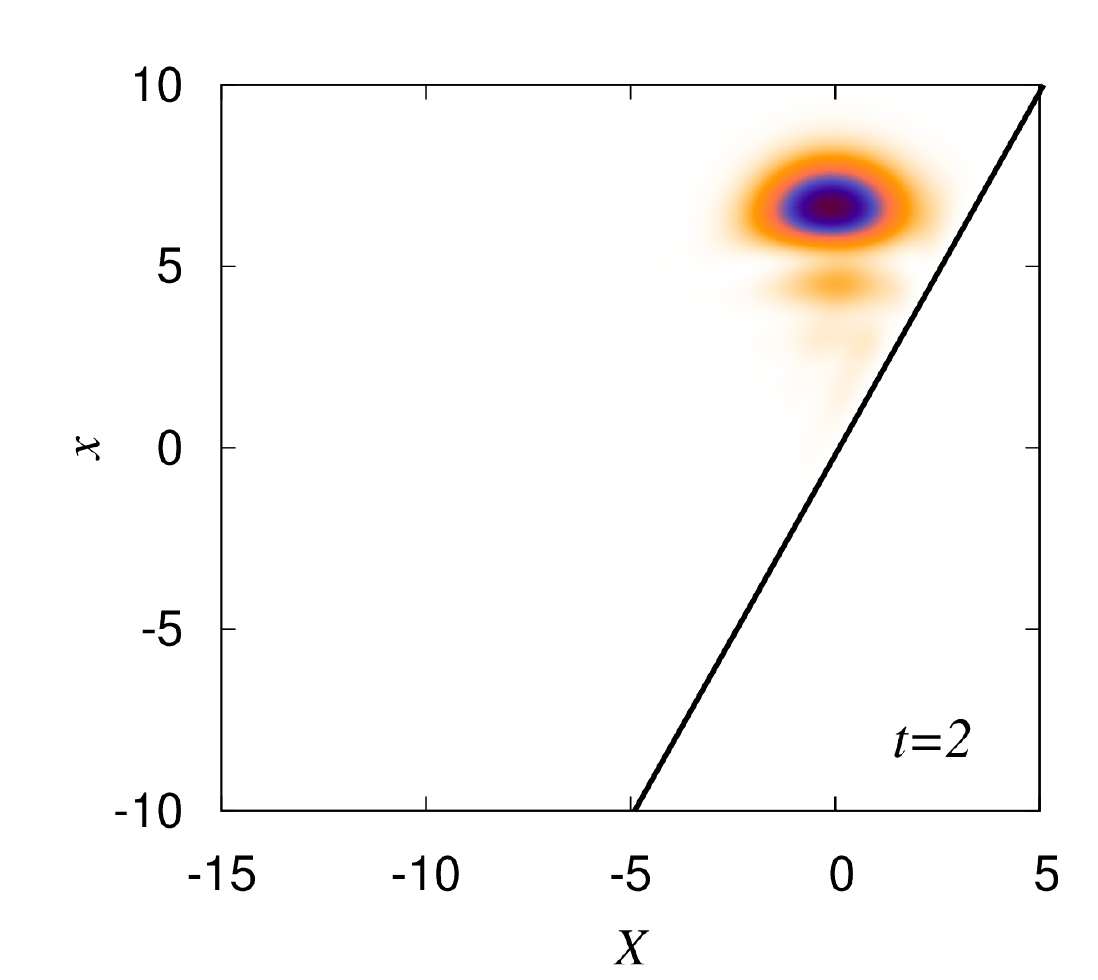} \tabularnewline
\includegraphics[width=0.475\linewidth]{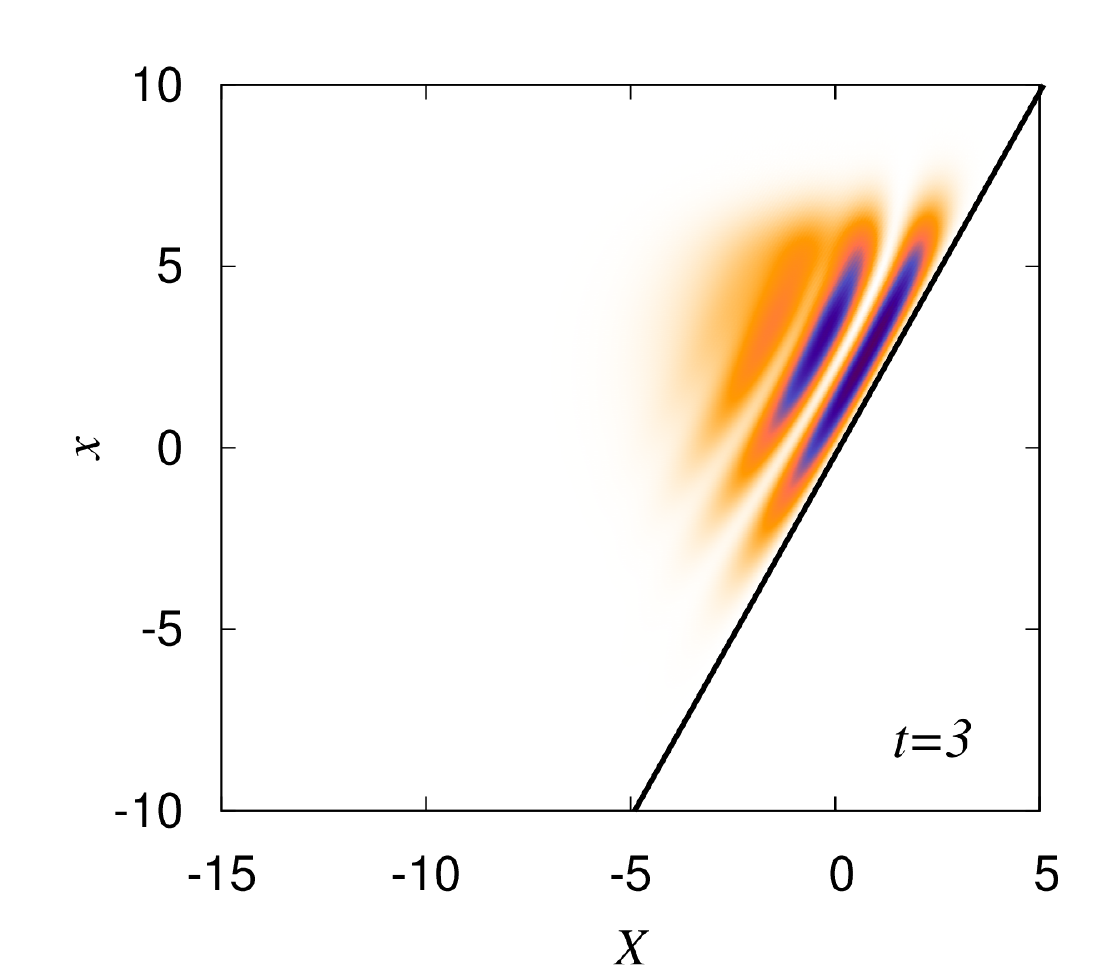} & \includegraphics[width=0.475\linewidth]{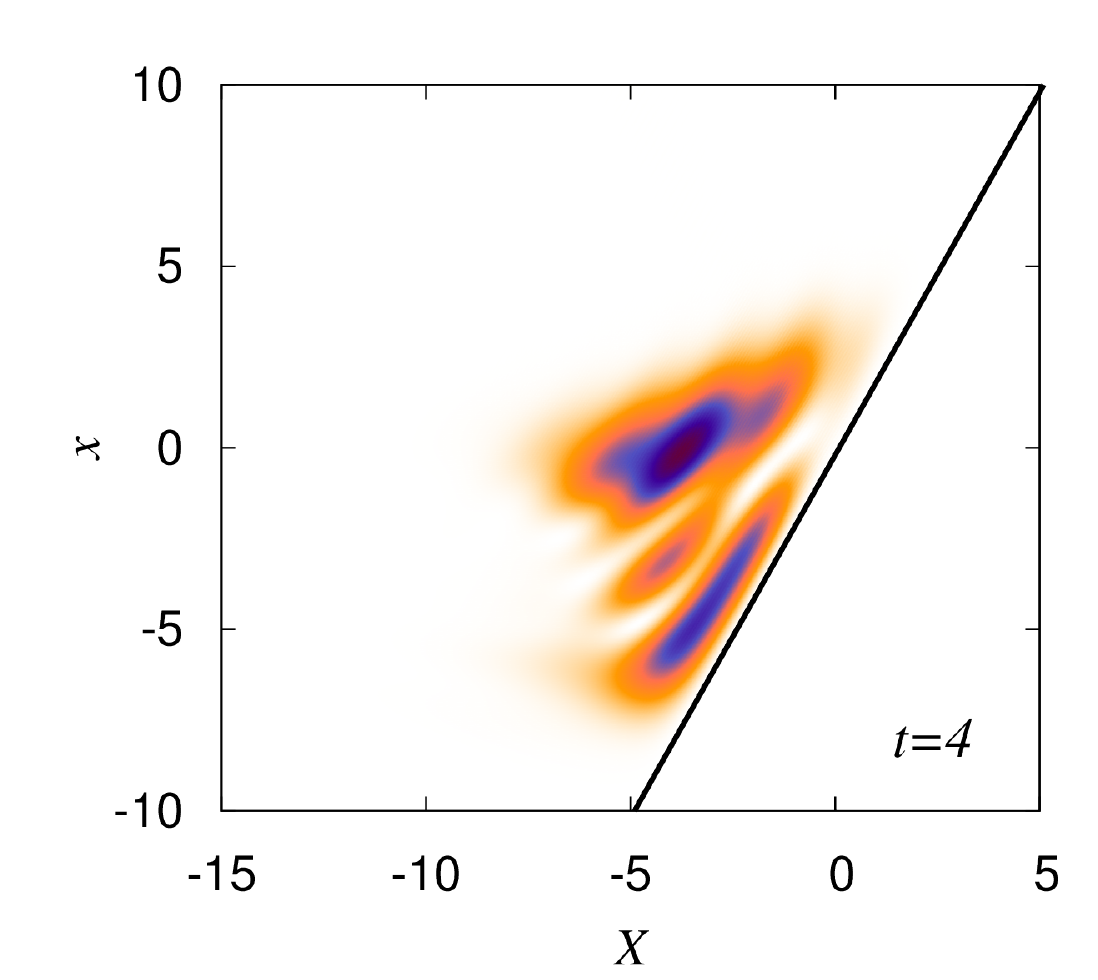} \tabularnewline
\end{tabular}
\caption{
Four panes, similar to those in Fig.~\ref{fig:1dtime}, show the wave function $|\Phi(X,x)|^2$ at most representative moments of time $t=1,2,3$ and 4, 
during the high energy collision with the impenetrable wall. Here  $K_0=5,$  the 
remaining parameters being the same as in Fig.~\ref{fig:1dtime}.
\label{fig:1dtimehigh}
}
\end{figure}
Let us return to the problem of scattering illustrated in Fig.~\ref{fig:Schematic-picture}. The time dependent picture of the 
scattering process is shown in Fig.~\ref{fig:1dtime} with a series of four plots showing the two-dimensional wave function using a density plot
at four different times.
The plot of the 
density projection onto the center of mass coordinate $X,$ which is the time dependent analog of Fig. \ref{fig:osc_wallnorm}, is shown 
below each of the four snapshots. 
The initial wave function at $t=0,$ shown on the first panel, is selected as the ground state wave function for the intrinsic potential, and
as a moving Gaussian wave packet for the center of mass coordinate, 
\begin{equation}
\Phi(X,x)=  \frac{1}{\sqrt{\sigma_0\sqrt{\pi}}}\exp\left[\frac{1}{2\sigma_0^2}(X-X_0)^2+ i K_0 X \right ]\,\psi_0(x). 
\label{eq:istate}
\end{equation}
In this example $\sigma_0=2,$ $X_0=-5,$ and initial momentum $K_0=1,$ all quantities are being expressed here 
in dimensionless units of distance as defined earlier. While this time dependent consideration is different from the stationary state formulation 
studied above, the series of snapshots for different times shown in Fig.~\ref{fig:1dtime} highlights some similar features.  

At high energies the dynamics of virtual excitations is complex; 
this is illustrated in Fig.~\ref{fig:1dtimehigh}, where the initial wave packet is selected to have $K_0=5.$
Some semiclassical interpretation can be given to the stages of the process. Initial compression at $t=1$ is 
followed by two particles bouncing apart at $t=2.$ Having equal masses, their center of mass remains at the origin but the relative separation 
$x$ becomes large so that the particles are positioned roughly symmetrically on the opposite sides of the wall. Next at $t=3$ the
center of mass moves into $X<0$ region pressing  the interacting particle against the wall. Finally, the system is reflected at $t=4$
with the initial wave packet being considerably distorted.   

In comparison to the projection and VPM techniques discussed earlier, the time-dependent approach is substantially faster numerically; 
moreover, any potential $U(x_1,x_2)$ can be considered with ease in this approach. 
One has to keep in mind, however, that it is not always easy to provide quantitative answers to stationary state questions, such as 
determination of scattering phase shifts in this example, using time-dependent techniques. 
The exact choice of the initial 
state as well as the energy uncertainty of the initial state can be important for some stages of time evolution.

The physics of decay of unstable states represents a particularly important class of time-dependent process.
The familiar exponential decay law is only an incomplete picture, 
requiring some subtle approximations, and being valid only within certain time limits. 
The complex intrinsic dynamics that can occur in the decaying many-body system further complicates the time evolution. 
The non-exponential decay laws in quantum mechanics have been studied and revisited by many authors (
see Ref.~\cite{Peshkin:2014} and references therein).  
The presence of three regimes, namely, initial, exponential, and long-time power law, appears to be a universal feature
of the decay processes.  The transitions from one regime to another are accompanied by the interference of corresponding
quantum amplitudes that is seen as oscillations on the decay curve.

\begin{figure}[!ht]
\begin{tabular}{cc}
\includegraphics[width=0.475\linewidth]{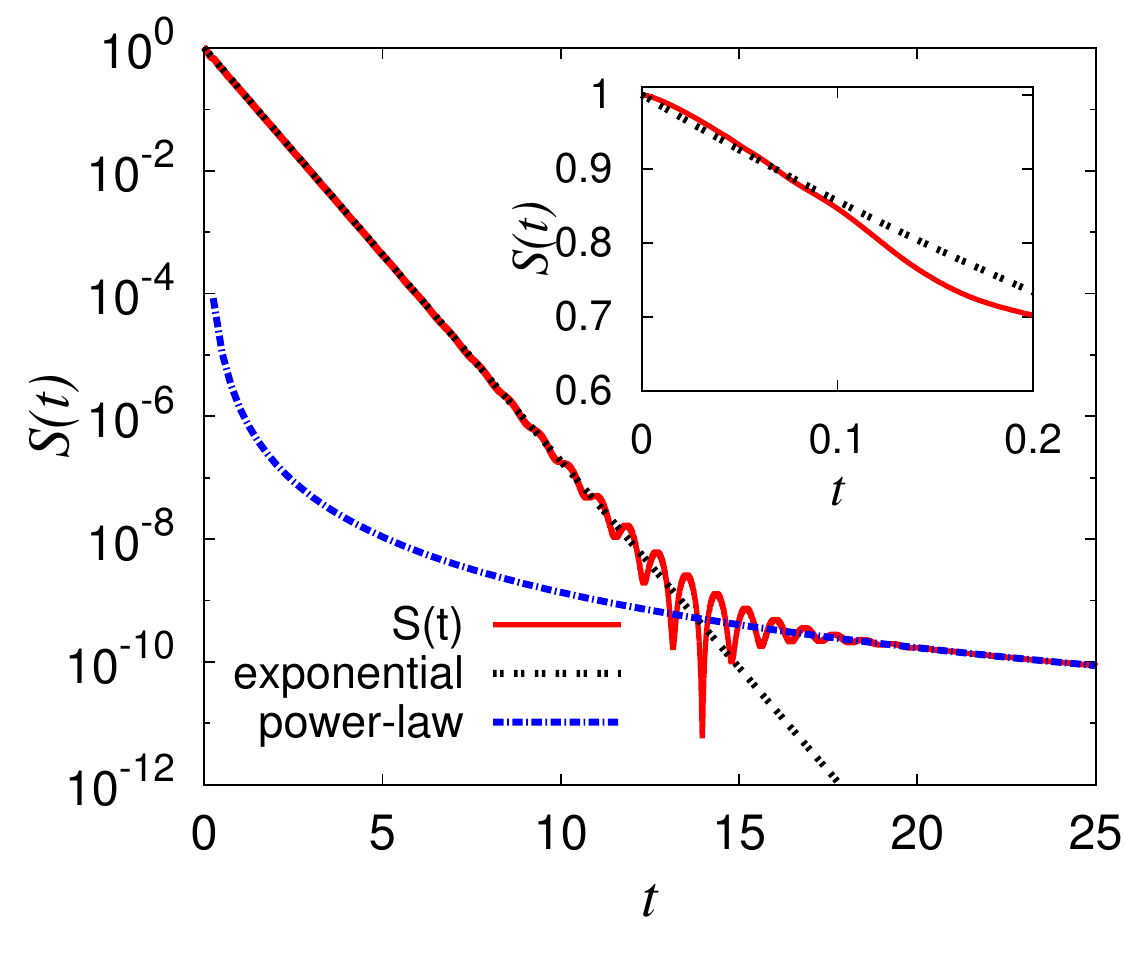} &  \includegraphics[width=0.475\linewidth]{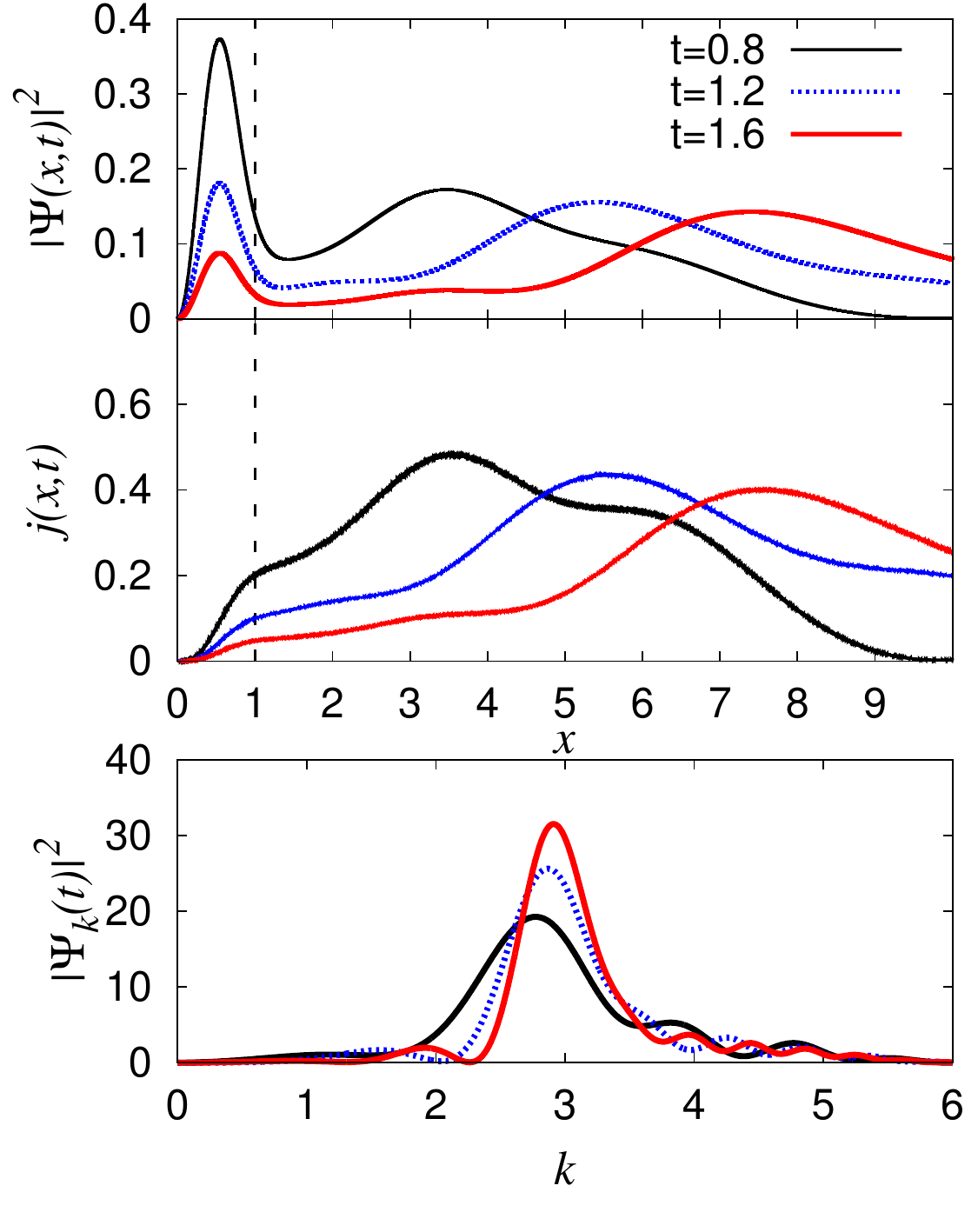}\tabularnewline
\end{tabular}
\caption{
Left: 
survival probability $S(t)=|\langle \Psi(0)|\Psi(t) \rangle |^2$ is shown as a function of time (solid red line). 
The exponential decay law, where mean lifetime $\tau=0.65$ is known from the poles of the scattering matrix, 
is shown with a double-dotted black line, the background 
component that decays following a power law is shown with a dot-dash  blue line.  The survival probability at very early times is shown in inset.
Right: wave function of a decaying state is shown at times  $t=0.8$, 1.2, and 1.6:
upper panel shows the probability distribution $|\Psi(x)|^2$,  middle panel displays current $j(x,t),$  and
 the wave function in momentum space is shown in the lower panel. Here the strength of the delta function $G=6,$ in units 
 where $\hbar=2m=1.$
\label{fig:winter}
}
\end{figure}

As another demonstration of the time-dependent technique based on the Trotter-Suzuki expansion and as an introduction to the 
section that follows,  we demonstrate in Fig.~\ref{fig:winter} the decay process in Winter's model \cite{Winter:1961}, which has been a very 
popular tool for exploring non-exponential features in decays. In this model a particle is confined to a region $x\ge0$ by an 
impenetrable wall at $x=0$ and is held by a delta barrier at $x=1.$ The initial state at $t=0$ is taken as $\Psi(x,0)=\sqrt{2}\sin(\pi x).$
The survival probability shown in the solid red line on the left panel of Fig.~\ref{fig:winter} 
illustrates the three general regimes: pre-exponential, exponential, and post-exponential. 
Oscillations can be seen in transitional regions. The snapshots of the wave function at different times are shown on the right. 

The pre-exponential behavior at very early times is influenced by the memory of how the state was created and, in particular, by the high 
energy components in the state. Later in time the internal structure and transitions between the intrinsic states become relevant. 
Short times correspond to remote energy components where the presence of other resonant states is to be considered. 
The high energy components have much shorter lifetimes and decay quickly leading to an exponential decay phase. This phase is dominated 
by a single resonant component, the radiating state, 
so that the wave function retains its shape while decreasing in amplitude. This can be seen on the 
right panel of Fig.~\ref{fig:winter}. In the same figure one can also trace a moving away background component. 
The background contains very low energy particles; being far off-resonance, they essentially do not interact but move slowly away from the 
interaction region. Near the decay threshold 
the number of such particles with a certain energy is determined by the available phase space, which for neutral 
particles scales with energy following a power-law $E^{l+1/2} $ where $l$ is the angular momentum quantum number. 
This type of scaling leads to non-resonant
components that follow a power-law decay $S(t)\sim 1/t^{2\ell+3}.$ While the non-resonant component can be very small in the initial state,
eventually it becomes dominant due to its slower-than-exponential decay. Further discussion of decay processes in quantum mechanics and
other examples can be found in Ref. \cite{Peshkin:2014}. The near-threshold phase space scaling with energy 
which leads to power-law decay at remote 
times is an important consideration in the Time Dependent Continuum Shell Model approach that is discussed in the following section, see also 
Refs.~\cite{Volya:2005PRL,Volya:2006, Volya:2009}, as well as in more complicated sequential decay processes \cite{Volya:2012}.

\section{Time dependent continuum shell model\label{sec:tdcsm}}
\subsection{Continuum Shell Model}
A seamless transition between structure physics and reactions is one of the central present-day 
theoretical problems. The computational aspect associated 
with transitions from discrete levels to a continuum of reaction states is especially challenging.  
The Continuum Shell Model approach \cite{Volya:2005PRL,Volya:2006} 
and  its time-dependent version, in particular, is one among several theoretical tools confronting these issues. 
In the Continuum Shell Model  the 
Feshbach projection formalism \cite{Feshbach:1958,Mahaux:1969} 
is used 
to express the exact dynamics in the full Hilbert space using  
an effective Hamiltonian in the projected intrinsic subspace of interest, ${\cal Q}:$
\begin{equation}
{\cal H}(E)=H_{{\cal QQ}}+\tilde{H}(E) \quad {\rm where}\quad \tilde{H}(E)=H_{{\cal QP}}\,\frac{1}{E-H_{{\cal PP}}}\,H_{{\cal PQ}}.
                                                                     \label{eq:heff1}
\end{equation} 
Here the effective Hamiltonian contains $H_{{\cal QQ}}$ which is the part of the original Hamiltonian that acts in the space ${\cal Q},$
and the energy-dependent non-Hermitian term $\tilde{H}(E),$ that emerges from the coupling of the space ${\cal Q}$ to an external space
containing a continuum of reaction states, ${\cal P}.$ 

In practical applications the intrinsic space ${\cal Q}$ is assumed to represent the configuration space of the traditional shell model, 
built from states $|1\rangle$ that are Slater determinants constructed from bound-state single-particle wave functions.
The space ${\cal P}$ contains continua of reaction states $|c,E\rangle$ characterized by the channel index, $c$, and 
the continuous  energy parameter $E$.  There is a certain threshold energy $E^{(c)}_{\rm thr}$ for each channel $c.$  
The energy-dependent non-Hermitian effective Hamiltonian (\ref{eq:heff1}) is then represented by a matrix 
${\cal H}_{12}(E)\equiv \langle 1 |{\cal H}(E) |2 \rangle,$
\begin{equation}
{\cal H}_{12}(E)=H_{12}+\Delta_{12}(E)-\frac{i}{2}W_{12}(E)\,,\quad {\rm where}
\label{eq:heff}
\end{equation} 
$$
 \Delta_{12}(E)=\sum_c {\rm PV}\int_{E^{(c)}_{\rm thr.}}^\infty dE' \frac{A^c_1(E') {A^c_2}^*(E')}{E-E'},\,\,\,  W_{12}(E)=2\pi\sum_{c {\rm (open)}} {A^c_1(E) {A^c_2}^*(E)},
$$
and the channel amplitudes are the matrix elements $A^c_1(E)=\langle 1|H|c,E\rangle.$ 
The traditional shell model Hamiltonian is recovered when the internal space ${\cal Q}$  is isolated and thus is decoupled, $A^c_1(E)=0.$

The computational challenges of the traditional shell model approach are well known, they are mainly associated with the need to find 
some selected eigenvalues and eigenvectors of the Hamiltonian matrix $H_{12}.$ The matrix is generally sparse, thanks to few-body nature of the underlying nucleon-nucleon interactions which inhibits mixing of very remote configurations, thus iterative techniques such as Lanczos approach 
are commonly used. 

The physics of weakly-bound and unstable nuclear systems is much more rich as questions of interest span from properties of bound states to features 
in scattering cross sections. 
Narrow resonances are well characterized by the usual properties of bound states with the decay width being an additional characteristic. 
This requires the non-Hermitian eigenvalue problem ${\cal H}(E)|{\cal I}\rangle={\cal E} |{\cal I}\rangle$ to be solved. 
The resulting complex energies ${\cal E}$ represent positions of resonances, 
$E= {\rm Re}({\cal E})$, and their widths, $\Gamma=-2\,{\rm Im}({\cal E}).$
The most practical technique here is to start with the perturbative treatment and evaluate the term $\tilde{H}(E)$ associated with continuum, 
using the wave functions of the traditional shell model Hamiltonian $H_{{\cal QQ}}.$ As coupling to the continuum increases the states become broad and one is forced to treat the non-Hermitian energy-dependent eigenvalue problem as an iterative non-Hermitian diagonalization process.   
In this limit a major problem is associated with the physical interpretation of the resonances and their widths.

Formally, the energy-dependent non-Herminitan Hamiltonian provides an exact propagator for the intrinsic space and therefore the 
scattering matrix  is 
$$
{\bf S}_{cc'}(E)=\exp(i\xi_c+i\xi_{c'})\,\left [\delta_{cc'} -2\pi i  {\bf T}_{cc'}(E) \right ], \quad {\rm where} 
$$
\begin{equation} 
{\bf T}_{cc'}(E)= \sum_{12}
A^{c}_1(E) \left \{\frac{1}{E-{\cal H}} \right \}_{12} A_2^{c'}(E).
\label{eq:cross}
\end{equation}
Here $\xi_{c}$ is a potential (direct-reaction) phase. 
The matrix is unitary (see Ref~\cite{Volya:2009}) and the unitarity is related to a factorized form of the imaginary  $W_{12}(E)$ in Eq.~(\ref{eq:heff}). 
The eigenvalues of the non-Hermitian Hamiltonian are therefore poles of the scattering matrix. 
In the limit of broad resonances one has to address the reaction problem where obtaining a reaction cross section is the main goal. 
There are several numerical challenges associated with  Eq.~(\ref{eq:cross}), many of these challenges being similar to the ones discussed
in Sec.~\ref{sec:proj}. First, the size 
of the Hamiltonian matrix and the complex arithmetic involved are not making this problem simpler as compared to matrix diagonalization. 
Second, the scattering energy $E$ represents a running parameter so that the procedure should be repeated for all energies of interest. Finally, 
the problem is numerically unstable: bound states, as well as resonances with widths ranging by many orders of magnitude, may be encountered 
and should be treated consistently. 
All of these technical issues are resolved by the Time-dependent Continuum Shell Model approach 
which we discuss next. 

\subsection{Time-dependent many-body evolution operator}
The many-body wave
function follows the time evolution which is a Fourier image of the retarded propagator involved in the scattering matrix (\ref{eq:cross}):
\begin{equation}
{G}(E)=\frac{1}{E-H}=-i\int_{0}^{\infty}dt\,\exp(iEt)\exp(-iHt).
\label{eq:Fourier}
\end{equation}
Here $H$ is an arbitrary Hamiltonian, but as discussed below, it is advantageous to include a factorized imaginary part $W$ 
using a different procedure
described in Sec.~\ref{sec:dyson}. 
Thus, we view $H$ as being a Hermitian Hamiltonian of the traditional shell model in which case it is set to have an infinitesimal negative-definite imaginary part. 
The time-dependent evolution operator can be factorized using a Chebyshev polynomial expansion method, 
see Ref.~\cite{Loh:2001,Ikegami:2002,Volya:2009}.
 \begin{equation}
\exp(-iHt)=\sum_{n=0}^{\infty}\,(-i)^{n}(2-\delta_{n0})\, J_{n}(t)\, T_{n}(H),\label{eq:EVOL}\end{equation}
where $J_{n}$ is the Bessel function of the first kind and $T_{n}$ represents 
Chebyshev polynomials. 
The Chebyshev polynomials, defined as
$T_{n}[\cos(\theta)]=\cos(n\theta)$
or, in explicit form,
\begin{equation}
T_{n}(x)=\frac{n}{2}\,\sum_{k=0,1,\dots}^{k\le n/2}\,\frac{(-1)^{k}}{n-k}\,\left(\begin{array}{c}
n-k\\
k\end{array}\right)(2x)^{n-2k},\end{equation}
provide a complete
set of orthogonal functions covering uniformly the interval [-1, 1]. In contrast, Taylor expansion relies on power functions which favor the 
edges of the interval and thus are more sensitive to extreme eigenvalues.  
The
``angular addition'' identity 
\begin{equation}
2T_{n}(x)T_{m}(x)=T_{n+m}(x)+T_{n-m}(x)\,,\quad n\ge m\label{eq:Tsum}\end{equation}
which follows from the definition, allows one to obtain these polynomials using the recurrence relation 
\begin{equation}
T_0(x)=1,\quad T_1(x)=x,\,\,\,{\rm and}\quad T_{n+1}(x)=2x T_n(x)-T_{n-1}(x).
\end{equation}
Therefore, the process of evaluation of Chebyshev polynomials of the Hamiltonian operator is an iterative procedure, similar to the one in 
Lanczos approach.  
For a given initial state $|\lambda\rangle\equiv|\lambda_{0}\rangle,$ a sequence $|\lambda_{n}\rangle=T_{n}(H)|\lambda\rangle$ can be 
constructed as
\begin{equation}
|\lambda_{0}\rangle=|\lambda\rangle,\quad|\lambda_{1}\rangle=H|\lambda\rangle,
\,\,\text{and}\,\,|\lambda_{n+1}\rangle=2H|\lambda_{n}\rangle-|\lambda_{n-1}\rangle.\label{eq:Titeration}\end{equation}
For overlap functions, assuming Hermitian $H,$ one can also use the following identity 
\begin{equation}
\langle\lambda'|T_{n+m}(H)|\lambda\rangle=2\langle\lambda'_{m}|\lambda_{n}\rangle-\langle\lambda'|\lambda_{n-m}\rangle,\quad n\ge m.\label{eq:double-angle}\end{equation}

Well controlled energy resolution is one advantage of the method.  
In applications of the method the energy interval $[E_{min},E_{max}],$ which should contain all eigenvalues of $H,$ is 
mapped onto $[-1,1]$ by rescaling the Hamiltonian as
$H\rightarrow(H-\overline{E})/{\Delta E}$ where 
$\overline{E}=(E_{max}+E_{min})/2$ and $\Delta E=(E_{max}-E_{min})/2.$ 
For a desired energy resolution $\Delta E/N$ where $N$ is some even integer number, the discrete Fourier transform allows one to 
evaluate Green's function in the corresponding energy points of the rescaled interval $E_{p}=p/N$ with
$p=-N/2\dots N/2,$ 
\begin{equation}
\langle\lambda'|G(E_{p})|\lambda\rangle=-i\pi\left\{ \sum_{\tau=0}^{N-1}\, e^{2\pi ip\tau/N}\sum_{n=0}^{n_{\text{max}}(\tau)}(-i)^{n}(2-\delta_{n0})J_{n}(\pi\tau)\langle\lambda'|T_{n}(H)|\lambda\rangle\right\} .
\end{equation}
This requires the evaluation of the evolution operator at times $t=\pi\tau,$ where $\tau=0\dots N-1.$ 
For each desired time point $\tau$ the number of terms in expansion (\ref{eq:EVOL}) needed for convergence is denoted as $n_{\text{max}}(\tau).$
The asymptotic of Bessel functions $J_{n}(x)\approx\sqrt{1/(2\pi n)}[ex/(2n)]^{n}$
suggests $n_{\text{max}}(\tau)\approx e\pi\tau/2\approx 4\tau.$ 
At fixed values of $n$
but for large times the convergence remains stable due to 
 $J_{n}(t)\approx\sqrt{2/(\pi t)}\cos(t-\pi n/2-\pi/4)$ in this limit. 
 For the desired energy resolution $\Delta E/N$ the propagation in time has to be extended up to $\approx\tau N$ which requires 
 $n_{\text{max}}\approx 4N;$ therefore $2N$ matrix-vector multiplications are required if one also uses Eq. (\ref{eq:double-angle}).

The time-dependent approach provides the Green's function for all energies at once; it is also exceptionally stable numerically when 
dealing with very narrow resonances or with stable states. Indeed, the time-dependent behavior of stationary states is regular and the 
corresponding delta function in energy is well handled by Fourier transform, which at the desired energy resolution properly conservers the integrated strength. 

\begin{figure}[!ht]
\begin{centering}
\begin{tabular}{cc}
\includegraphics[width=0.475\linewidth]{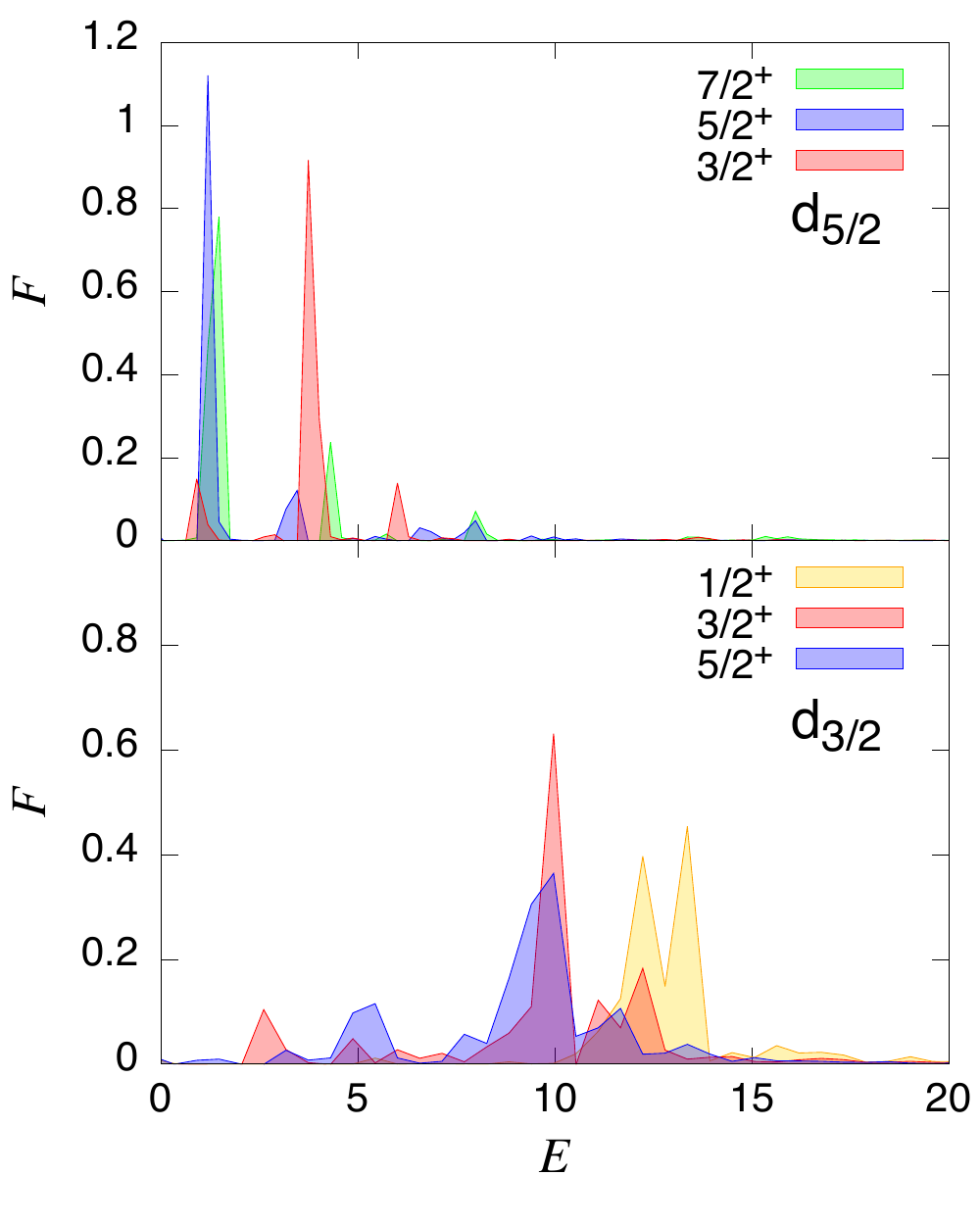} & \includegraphics[width=0.475\linewidth]{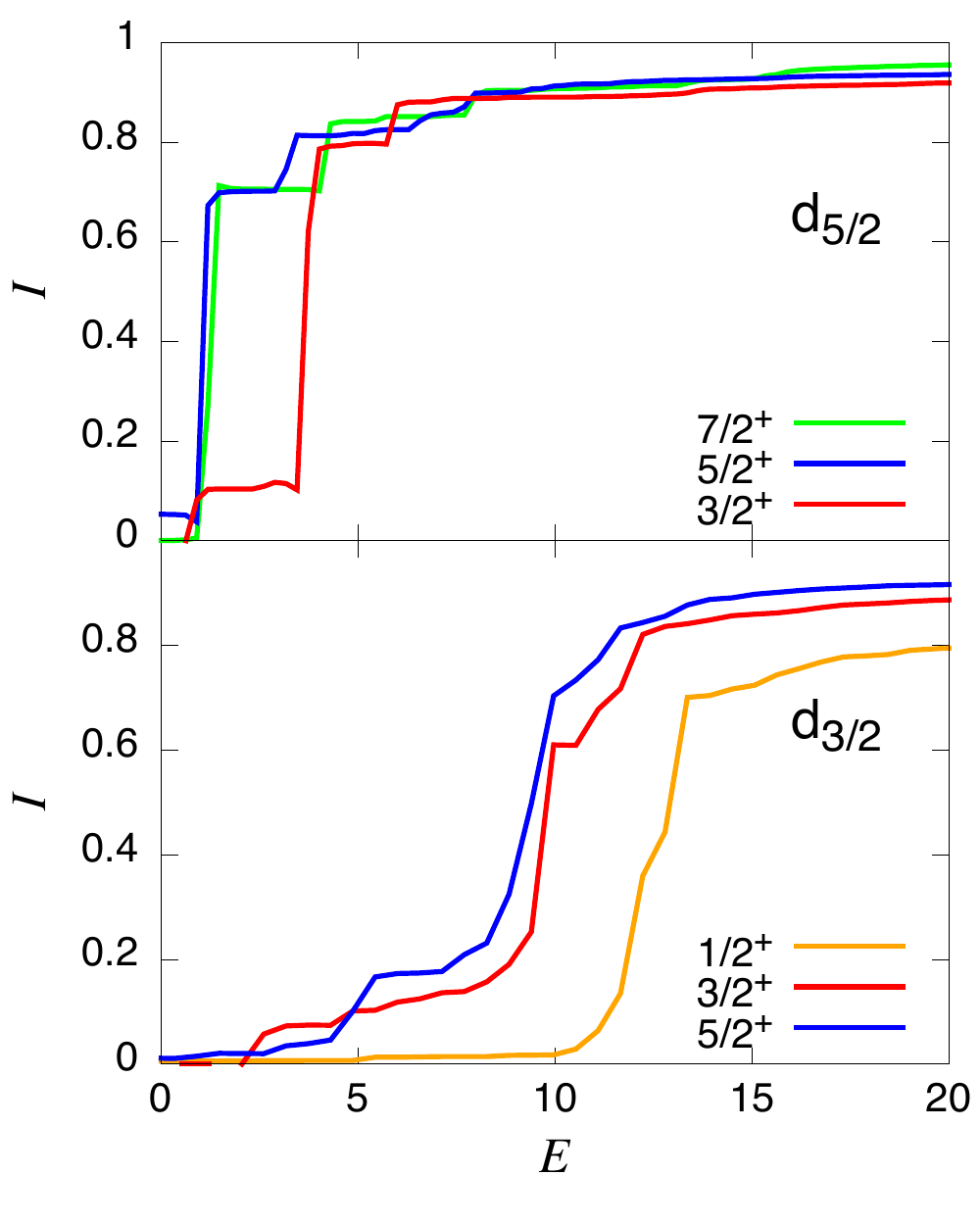} \tabularnewline
\end{tabular}
\end{centering}
\caption{Single-particle strength function (left) and cumulative or integrated strength function (right) are shown as functions of excitation energy 
(in units of MeV) for  $^{15}$N.
\label{fig:strength}
}
\end{figure}

In order to illustrate the approach, let us consider strength and integrated strength functions  
defined for a given state $|\lambda\rangle$  as
\begin{equation}
F_{\lambda}(E)=\langle\lambda|\delta(E-H)|\lambda\rangle=-\frac{1}{\pi}\,{\rm Im}\,\langle\lambda|G(E)|\lambda\rangle,\quad
I_\lambda(E)=\int_{-\infty}^EF_\lambda(E')dE'\,.
\label{eq:strength-function}\end{equation}
In Fig.~\ref{fig:strength} both strength (left) and integrated strength (right) functions are shown for $^{15}$N for  
neutron channels where $|\lambda\rangle$ corresponds to different angular momentum channels constructed from $1^{+}$ ground state in $^{14}$N coupled to
a single nucleon on either $d_{5/2}$ (top panels) or $d_{3/2}$ (bottom panels) single-particle states. This theoretical study follows recent 
experimental work in Ref.~\cite{Mertin:2014}. The full $p$-$sd$ valence space is used with the Hamiltonian from Ref. \cite{Utsuno:2011}. 
With about 
$10^7$ $m$-scheme basis states, obtaining and computing strength functions in energy regions around 20 MeV of excitation is impractical;  the time-dependent method provides an excellent alternative. 

\subsection{Sherman-Morrison-Woodbury relations\label{sec:dyson}}
It is certainly possible to implement the Chebyshev polynomial expansion procedure for a full non-Hermitian Hamiltonian using 
Eq.~(\ref{eq:EVOL}); however the factorized structure of $\tilde{H}$ offers 
a different alternative which is much more computationally advantageous. 
The two propagators corresponding to Eq. (\ref{eq:heff1})
\begin{equation}
G(E)=\frac{1}{E-H_{{\cal QQ}}}\quad{\rm and}\quad{\cal G}(E)=\frac{1}{E-{\cal H}(E)}
\label{eq:GHV}
\end{equation}
can be related  through Dyson's equation ${\cal G}(E)=G(E)+G(E)\tilde{H}(E){\cal G}(E).$
Since the contribution from the continuum emerges in the factorized form 
\begin{equation}
\tilde{H}(E)=\sum_{cc'}|c\rangle\tilde{\bf H}_{cc'}(E)\langle c'|,
\end{equation}
the expression for the full propagator can be found in a closed form in the space spanned by the channel states
\begin{equation}
\mathbb{G}={\bf G}\left [{\bf 1}-\tilde{\bf H}{\bf G}\right ]^{-1}=\left [ {\bf 1}-{\bf G}\tilde{\bf H}\right ]^{-1}{\bf G}.\label{eq:G}
\end{equation}
The operators here are represented by matrices
in the channel subspace $\mathbb{G}_{ab}=\langle a|\mathcal{G}(E)|b\rangle$
and ${\bf G}_{ab}=\langle a|G(E)|b\rangle$. In computer science these relations are known as Sherman-Morrison-Woodbury matrix
inversion equations~\cite{Press:2002}. The unitarity of the scattering matrix immediately follows from these relations, see \cite{Volya:2009}.

\begin{figure}[!ht]
\begin{center}
\includegraphics[width=9cm]{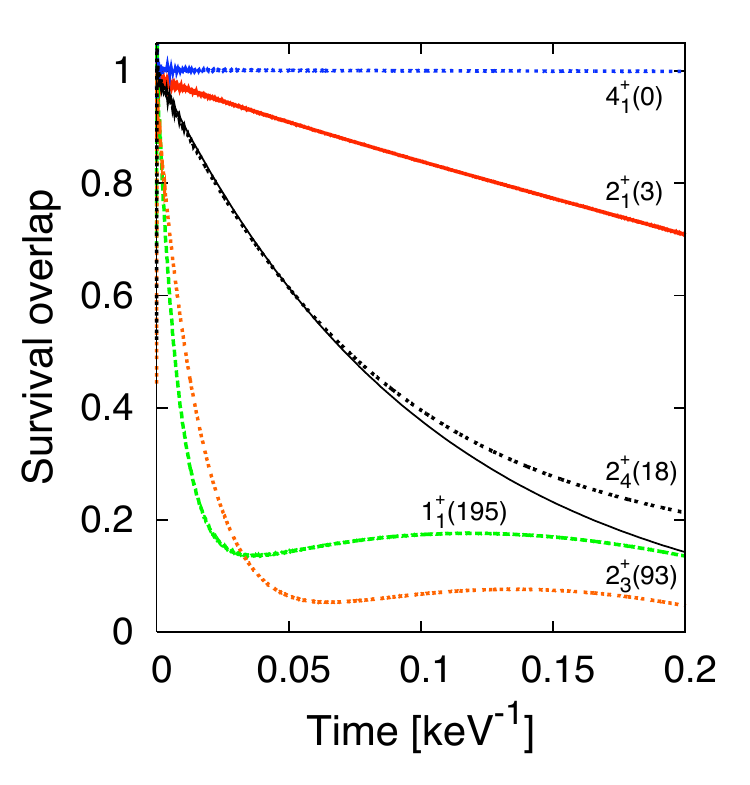}
\end{center}
\caption{Time evolution of several low-lying states in $^{24}$O. The absolute
value of the survival overlap $\left|\langle\alpha|\exp(-i{\cal H}t)|\alpha\rangle\right|$
is shown as a function of time. Different lines, as marked, correspond
to states $\alpha(E_\alpha,\Gamma_\alpha):$ $2_{1}^{+}$ (4.180, 2.7), $1_{1}^{+}$
(5291, 195.1), $4_{1}^{+}$ (6947, 0.0), $2_{3}^{+}$ (8107, 92.5) and
$2_{4}^{+}$ (9673, 17.5). They are eigenstates of the traditional
USD SM but are non-stationary resonances in the TDCSM, except
for the $4_{1}^{+}$ state which due to its high spin does not decay within
the $sd$ valence space. To emphasize the non-exponentiality in the
decay law the unmarked solid line shows the $\exp(-\Gamma_\alpha t/2)$ function
with parameters for the $2_{4}^{+}$ state. \label{fig:time}}
\end{figure}

\begin{figure}[!ht]
\begin{minipage}[h]{0.65\textwidth}
\includegraphics[width=8cm]{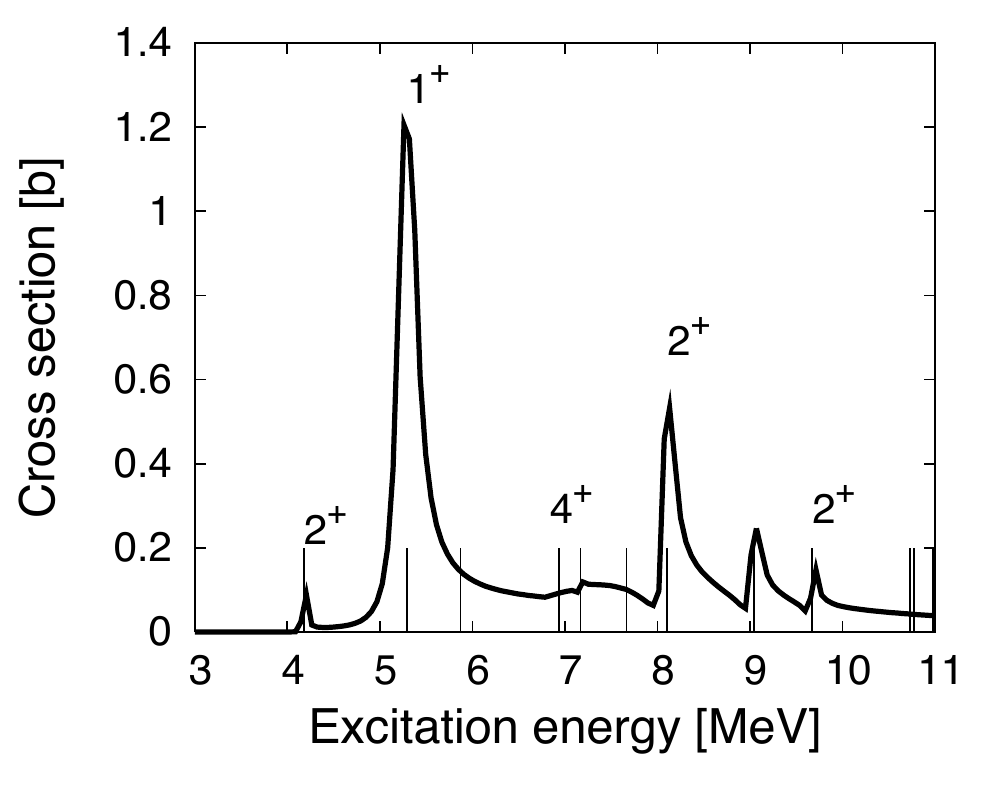}
\end{minipage}
\begin{minipage}[h]{0.3\textwidth}
\caption{Scattering cross section for $^{23}$O$(n,n)$$^{23}$O reaction showing resonances in $^{24}$O.\label{fig:time1}}
\end{minipage}
\end{figure}

We illustrate the TDCSM approach in its complete form in Figs.~\ref{fig:time} and ~\ref{fig:time1} where the resonances in $^{24}$O are considered. 
The system is treated in the $sd$ valence space using the USD shell model Hamiltonian \cite{Brown:1988HB}.
In Fig.~\ref{fig:time} the norm of the survival amplitude is shown as a function of time for the following set of most representative states
$2_{1}^{+}$ (4180, 2.7), $1_{1}^{+}$
(5291, 195.1), $4_{1}^{+}$ (6947, 0.0), $2_{3}^{+}$ (8107, 92.5), and
$2_{4}^{+}$ (9673, 17.5).  The states are listed here with their excitation
energies followed by the decay widths, both in keV.  The initial wave functions at $t=0$ are taken as  eigenstates of the traditional shell model. 
For the states such as $4_{1}^{+}$, which cannot decay in this model due to 
high angular momentum, the norm of the survival amplitude remains constant. Narrow states, exhibit a nearly exponential decay, for the state
$2^+_4$ the survival amplitude expected in exponential decay is shown. The decay is non-exponential for broad states such as
$1_{1}^{+}$ and $2_{3}^{+}.$
In Fig.~\ref{fig:time1} the scattering cross section is shown for elastic neutron scattering on the ground state of $^{23}$O, 
where the same resonant states can be observed. 

The time-dependent approach provides an effective computational strategy for 
treating many-body systems that feature both bound and unbound states. 
In contrast to the stationary state formalism, the time dependent approach addresses the evolution of states in a natural way, 
thus providing a computationally robust and stable strategy, where experimental observables are easily recovered and fundamental 
principles of quantum mechanics, such as linearity and unitarity, are followed.
From the computational perspective,   
the matrix-vector multiplication, the most efficient operation available, is utilized in building the time evolution operator with full control of 
the desired energy and time resolution.  The specifics of the terms that emerge due to coupling to continuum in Feshbach projection formalism 
can be used to build the full evolution operator using Sherman-Morrison-Woodbury relations. 
TDCSM found broad practical applications, see Refs.~\cite{Mitchell:2013,Mitchell:2010,Rogachev:2007} for example.

\section{Conclusions}
As our interests shift towards open, reacting, decaying, and otherwise evolving quantum many-body systems, new theoretical and computational techniques must be developed to address multiple new challenges that emerge. The goal of this presentation is to highlight some of the methods used in the recent scientific projects. We use a simple model to demonstrate three distinctly different techniques. 
The most straight-forward method involves projecting the dynamics onto a set of basis states, allowing subsequently for the well-developed methods of linear algebra to be used; in certain reaction problems this method appears to have significant drawbacks associated with numerical instabilities and poor convergence. We demonstrate the Variable Phase Method that can treat reaction problems efficiently in a discretized coordinate space.
Finally, we consider explicitly time-dependent techniques that are perhaps most adequate for the time-dependent dynamics associated with decay. 
We put forward the Time Dependent Continuum Shell Model approach, as a practical tool and demonstrate its application to realistic problems 
in nuclear physics.

This material is based upon work
supported by the U.S. Department of Energy Office of Science, Office of Nuclear Physics
under Award Number DE-SC0009883. The author is grateful to N. Ahsan, M. Peshkin, and V. Zelevinsky for collaboration.


\end{document}